\documentclass[twocolumn,showpacs,superscriptaddress,amssymb,10pt,pra,floatfix,aps,longbibliography]{revtex4-2}
\usepackage[dvips]{graphicx}

\usepackage{longtable}
\usepackage{dcolumn}
\usepackage[dvips]{graphicx}
\usepackage{bm}
\usepackage{lipsum}
\usepackage{bbm}
\usepackage{color}

\usepackage{times}
\usepackage{nicefrac}
\usepackage{amsmath}
\usepackage{amsfonts}
\usepackage{amssymb}
\usepackage{amsthm}
\usepackage[normalem]{ulem}
\newcolumntype{.}{D{x}{}{-1}}
\usepackage{orcidlink}
\usepackage{epsfig}
\usepackage{leftidx,amsmath}
\usepackage{natbib} 
\usepackage[utf8]{inputenc}
\usepackage{braket}
\usepackage{makecell}
\usepackage[flushleft]{threeparttable}
\usepackage{multirow}
\usepackage{isotope}
\usepackage{array}
\usepackage{siunitx}

\begin{document}
%%%%%%%%%%%%%%%%%%%%%%%%%%%%%%%%%%%%%%%%%%%%%%%%%%%%%%%%%%%%%%%%%%%%%%%%%%%%%%%

\title{Wichmann-Kroll Correction to the Interelectronic Interaction in He- and Li-Like Ions}

\author{J.~Sommerfeldt, \orcidlink{0000-0002-3471-7494}}
\thanks{joso@mpi-hd.mpg.de}
\affiliation{Max Planck Institute for Nuclear Physics, D-69117 Heidelberg, Germany}

\author{V.~A.~Yerokhin, \orcidlink{0000-0002-2328-8444}}
\affiliation{Max Planck Institute for Nuclear Physics, D-69117 Heidelberg, Germany}

\author{Z.~Harman, \orcidlink{0009-0006-0554-1537}}
\affiliation{Max Planck Institute for Nuclear Physics, D-69117 Heidelberg, Germany}

\author{C.~H.~Keitel, \orcidlink{0000-0002-1984-1470}}
\affiliation{Max Planck Institute for Nuclear Physics, D-69117 Heidelberg, Germany}

\date{\today \\[0.3cm]}

\begin{abstract}
We present a theoretical study of the higher-order QED contribution to the interelectronic interaction in He- and Li-like ions, where a virtual electron-positron loop is inserted into the photon line of the one-photon exchange diagram. Our approach is based on the Dirac-Coulomb Green's function and accounts for the interaction of the virtual $e^+e^-$ pair with the electric field of the nucleus to all orders in $\alpha Z$, with $\alpha$ being the fine-structure constant and $Z$ the atomic charge number. We show that the numerical convergence of the involved integrals can be significantly improved by explicitly subtracting the non-gauge-invariant spurious contributions from the integrands. We present improved numerical values for this contribution to the Lamb shift over a wide range of nuclear charge numbers $Z$. Our calculations agree well with previous results by Artemyev and co-workers~[Phys. Rev. A \textbf{56}, 3529 (1997); Phys. Rev. A \textbf{60}, 45 (1999)] for He-like ions, but we find a discrepancy in the Li-like case. Moreover, we calculate the finite nuclear size correction to this diagram, which can reduce its size by more than 5\% for heavy ions. The improved QED calculations not only decrease the uncertainty of theoretical predictions for the interelectronic interaction in few-electron ions but the methods could also be used in the future to improve calculations of closely related one-electron two-loop QED diagrams.
\end{abstract}
\maketitle
\section{Introduction}
Highly charged ions (HCIs) have attracted increasing attention in the last few decades due to the unique opportunities they provide to test fundamental physics at the precision frontier. The electric field strength experienced by the electrons bound in an HCI is among the strongest that can be probed in a laboratory, and the energy levels can be treated rigorously within the framework of quantum electrodynamics (QED), rendering high-precision tests of strong-field QED theory possible~\cite{mps1998, sbgk2018, kscs2018, ind2019}. Recent experiments have continuously reduced the uncertainty of Lamb shift and bound-electron $g$-factor measurements in heavy HCIs~\cite{bkms2003, gsbb2004, bctt2005, bbkc2007, uabd2017, bss2021, sdhh2022, lbds2024}. Alongside these experimental advances were many theoretical works that have performed increasingly sophisticated QED calculations motivated by the experimental progress~\cite{sap1998, asyp2005, pcjy2005, yas2015, ppv2017, mgka2021, sykh2025}. This task is very challenging for large nuclear charge numbers $Z$, because the expansion in the Coulomb interaction parameter $\alpha Z$, where $\alpha \approx 1/137$ is the fine-structure constant, does not converge well in this regime. Although the one-loop contributions to the all-order Lamb shift were computed several decades ago~\cite{raw1973, gyu1974, sam1988, moh1974, moh1982}, already at second order in $\alpha$ there are still contributions that are not known with sufficient precision and limit tests of strong-field QED in HCIs~\cite{yis2008, sdhh2022}.

\begin{figure}
    \centering
    \includegraphics[width=0.45\linewidth]{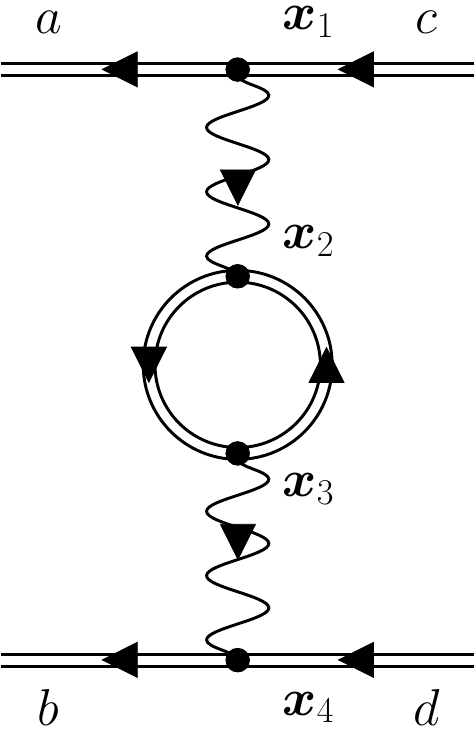}
    \caption{All-order in $\alpha Z$ Feynman diagram for the one-photon exchange correction with an inserted vacuum polarization loop for two electrons in initial states $c$, $d$ and final states $a$, $b$.}
    \label{Fig:Feynman}
\end{figure}
In this article, we present improved calculations of the vacuum polarization correction to the interelectronic interaction in few-electron ions, as described by the Feynman diagram in Fig.~\ref{Fig:Feynman}, where the double lines represent an electron or positron in the Coulomb field of the nucleus and the wavy lines denote virtual photons. This correction is relevant not only because it contributes to the two-electron QED part of the Lamb shift, but it is also closely related to the one-electron two-loop diagram, in which a virtual electron-positron loop is inserted into the photon line of the self-energy diagram (SVPE). The two-loop SVPE diagram is currently not known to all orders in $\alpha Z$ and is among the largest contributors to the theoretical uncertainty in HCIs~\cite{mks2023}. The diagram is closely related to Delbrück scattering since it can be interpreted as a Delbrück scattering of the virtual photon exchanged between the two electrons~\cite{mil1994, syss2023}. Moreover, in muonic atoms, the imaginary part of this diagram is connected to the pair production rate induced by bound-state transitions~\cite{aylv2026}.

The all-order correction shown in Fig.~\ref{Fig:Feynman} was first calculated by Artemyev and co-workers~\cite{asy1997, abps1999} and by Sunnergren~\cite{sun1998}. Although higher-order (in $\alpha Z$) corrections to this interelectronic effect were explicitly given in these studies, the reported numerical values had only limited accuracy and the influence of the finite nuclear size was neglected. In this work, we present improved calculations of the higher-order corrections to the two-electron QED diagram shown in Fig.~\ref{Fig:Feynman} for He- and Li-like ions. We develop a more efficient computational method by explicitly subtracting spurious terms that slow the numerical convergence. Moreover, we calculate finite nuclear size effects for different nuclear models and investigate the theoretical uncertainty of the computation.

The structure of this paper is as follows. In Section~\ref{Sec:Theory}, we present the basic formalism describing the vacuum polarization correction to the interelectronic interaction within the framework of bound-state QED including the expressions used to calculate the final contribution to the Lamb shift. The details of the computation are discussed in Section~\ref{Sec:Comp}, together with the remaining numerical uncertainty. Accurate numerical results are presented in Section~\ref{Sec:Res} for He- and Li-like ions over a wide range of nuclear charge numbers. Finally, the results are summarized in Section~\ref{Sec:Conc}. Relativistic units ($\hbar=m_e=c=1$) are used throughout this paper unless stated otherwise.

\section{Theoretical Framework} \label{Sec:Theory}
\subsection{Basic Formalism}
The total contribution to the Lamb shift represented by the Feynman diagram in Fig.~\ref{Fig:Feynman} for the interaction between two electrons in states $a$ and $b$ described by a Slater determinant wave function can be obtained using the standard Feynman rules of bound-state QED~\cite{sha2002} as
\begin{equation} \label{Eq:EVP1}
    E_\text{$e$VP$e$} = M(ab,ab;0) - M(ba,ab;\omega)~,
\end{equation}
where
\begin{widetext}
\begin{equation} \label{Eq:EVP}
\begin{aligned}
   M(ab,cd;\omega) = \frac{(4\pi\alpha)^2}{2\pi i} \int_{-\infty}^{+\infty}dz\int &d^3\boldsymbol{x}_1~d^3\boldsymbol{x}_2~ d^3\boldsymbol{x}_3~d^3\boldsymbol{x}_4~\psi^\dagger_a(\boldsymbol{x}_1)\alpha_\mu\psi_c(\boldsymbol{x}_1) D^{\mu\nu}(\omega,\boldsymbol{x}_{12})\\
    &\times\text{Tr}\left[\alpha_\nu G\left(\boldsymbol{x}_2, \boldsymbol{x}_3, z-\frac{\omega}{2}\right) \alpha_\rho G\left(\boldsymbol{x}_3, \boldsymbol{x}_2, z+\frac{\omega}{2}\right)\right] D^{\rho\sigma}(\omega,\boldsymbol{x}_{34}) \psi^\dagger_b(\boldsymbol{x}_4)\alpha_\sigma\psi_d(\boldsymbol{x}_4)~,
\end{aligned}
\end{equation}
\end{widetext}
and $\omega = E_a-E_b$ follows from energy conservation at each vertex. The correct sign of $\omega$ can be verified using the connection of the two-electron one-photon exchange diagram in Fig.~\ref{Fig:Feynman} to the three-electron two-photon exchange correction within the redefined vacuum approach, see Appendix~\ref{Ap:RedVac}. $D^{\mu\nu}(\omega,\boldsymbol{x}_{12})$ is the photon propagator, which, in Feynman gauge, can be written as
\begin{equation}
    D^{\mu\nu}(\omega,\boldsymbol{x}_{12})= g^{\mu\nu} \frac{\exp[i\sqrt{\omega^2+i\delta}\vert \boldsymbol{x}_{12}\vert]}{4\pi\vert \boldsymbol{x}_{12}\vert}~,
\end{equation}
where $\boldsymbol{x}_{12} = \boldsymbol{x}_{1}-\boldsymbol{x}_{2}$ and $\text{Im}[\sqrt{\omega^2+i\delta}]>0$ with $\delta$ being a small positive constant. The electron wave function $\psi_a$ in Eq.~\eqref{Eq:EVP} can be written as
\begin{equation}
    \psi_a(\boldsymbol{x}) = \left(\begin{array}{c}
         f^1_{n_a\kappa_a}(x)\chi_{\kappa_a}^{\mu_a}(\boldsymbol{\hat{x}}) \\
         if^2_{n_a\kappa_a}(x)\chi_{-\kappa_a}^{\mu_a}(\boldsymbol{\hat{x}})
    \end{array}\right)~,
\end{equation}
where $\chi_{\kappa_a}^{\mu_a}(\boldsymbol{\hat{x}})$ is the spin-angular spinor, and $f^1$ and $f^2$ are the large and small radial components, respectively. The Dirac-Coulomb Green's function $G(\boldsymbol{x}_1, \boldsymbol{x}_2, z)$ is given in its spectral representation as
\begin{equation}~\label{Eq:Green}
    G(\boldsymbol{x}_1, \boldsymbol{x}_2, z) = \sum_{n\kappa\mu} \frac{\psi_{n\kappa\mu}(\boldsymbol{x}_1)\psi^\dagger_{n\kappa\mu}(\boldsymbol{x}_2)}{E_{n\kappa}-z-i\delta}~.
\end{equation}
For a spherically symmetric potential, it is convenient to separate the radial part of the partial wave expansion of Eq.~\eqref{Eq:Green} as
\begin{equation} \label{Eq:RadG}
    G^{ij}_\kappa (x_1,x_2, z) = \sum_n \frac{f_{n\kappa}^i(x_1)f_{n\kappa}^j(x_2)}{E_{n\kappa}-z-i\delta}~,
\end{equation}
where the sum is to be understood as a summation over all bound states and an integration over the positive and negative continuum. The radial Green's function~\eqref{Eq:RadG} can be written as 
\begin{equation}
\begin{aligned}
G_\kappa(&r_2,r_1,z) = \frac{1}{w_\kappa(z)} \\
&\times\Bigg[\Theta(r_2-r_1) \left(\begin{array}{c}
F_{\kappa,\infty}^1 (r_2,z)\\
F_{\kappa,\infty}^2 (r_2,z)\\
\end{array}\right) \left(\begin{array}{c}
F_{\kappa,0}^1 (r_1,z)\\
F_{\kappa,0}^2 (r_1,z)\\
\end{array}\right)^T\\
&+\Theta(r_1-r_2) \left(\begin{array}{c}
F_{\kappa,0}^1 (r_2,z)\\
F_{\kappa,0}^2 (r_2,z)\\
\end{array}\right) \left(\begin{array}{c}
F_{\kappa,\infty}^1 (r_1,z)\\
F_{\kappa,\infty}^2 (r_1,z)\\
\end{array}\right)^T\Bigg]~,
\end{aligned}
\end{equation}
where
\begin{equation} \label{Eq:Wronskian}
w_\kappa(z) = r^2 [F^2_{\kappa,0}(r,z) F^1_{\kappa,\infty}(r,z)-F^1_{\kappa,0}(r,z) F^2_{\kappa,\infty}(r,z)]~,
\end{equation}
is the Wronskian, and $F^{1,2}_{\kappa,0}(r, z)$ and $F^{1,2}_{\kappa,\infty}(r, z)$ are the solutions of the homogeneous radial Dirac equation for a given energy value $z$
\begin{equation} \label{Eq:DiracRad}
\left(\begin{array}{cc}
1 + V(r) - z & -\frac{1}{r} \frac{\text{d}}{\text{d}r} r + \frac{\kappa}{r}\\
\frac{1}{r} \frac{\text{d}}{\text{d}r} r + \frac{\kappa}{r} & -1 + V(r) - z
\end{array}\right)\left(\begin{array}{c}
F_{\kappa}^1 (r,z)\\
F_{\kappa}^2 (r,z)\\
\end{array}\right)  = 0 ~,
\end{equation} 
that are regular at the origin and at infinity, respectively. For the Coulomb potential of a point-like nucleus, it is possible to find closed analytical solutions for Eq.~\eqref{Eq:DiracRad} in terms of Whittaker functions~\cite{mps1998, hyl1984}. Additionally, we consider realistic nuclear potentials to evaluate the influence of the finite nuclear size. In this case, we solve the radial Dirac equation numerically, see Refs.~\cite{sam1991,sap2026} for more details.

\subsection{Angular Integration}
To solve the angular integrals in Eq.~\eqref{Eq:EVP}, we use the multipole expansion of the photon propagator
\begin{equation}
    \begin{aligned}
        D^{\mu\nu}(\omega,\boldsymbol{x}_{12}) = g^{\mu\nu}\sum_{JM} &(-1)^M g_J(\omega, x_<, x_>)\\ &\times Y_J^M(\boldsymbol{\hat{x}}_1) Y_J^{-M}(\boldsymbol{\hat{x}}_2)~, 
    \end{aligned}
\end{equation}
where $ Y_L^M(\boldsymbol{\hat{x}})$ are spherical harmonics and the radial functions are
\begin{align}
    g_J(0, x_<, x_>) &= \frac{1}{2J+1} \frac{x_<^J}{x_>^{J+1}}\\
    g_J(\omega, x_<, x_>) &= i\omega j_J(\omega x_<) h^{(1)}_J(\omega x_>)~,
\end{align}
with $j_J(\omega x_<)$ and $ h^{(1)}_J(\omega x_>)$ being the spherical Bessel and Hankel functions, respectively.

Inserting the multipole expansion of the photon and electron propagators into Eq.~\eqref{Eq:EVP} and solving the angular integrals analytically~\cite{jbs1988}, one obtains
\begin{equation} \label{Eq:AmpFinal}
    \begin{aligned}
        &M(ab,cd;\omega) = \frac{\alpha^2}{2\pi i} \sum_{J_1 J_2,\kappa \kappa'} I_{J_1}(a,\kappa\mu;c,\kappa'\mu') \\ & \times I_{J_2}(\kappa'\mu',b;\kappa\mu,d)\int_{-\infty}^{+\infty} dz~R_{J_1}(a\kappa;c\kappa') R_{J_2}(\kappa'b;\kappa d)~,
    \end{aligned}
\end{equation}
with the angular part
\begin{equation} \label{Eq:AngInt}
    \begin{aligned}
        I_J(ab;cd) = &\sum_M(-1)^{j_a-\mu_a+j_b-\mu_b+J-M} \\
        &\times \left(\begin{array}{ccc}
            j_a & J & j_c \\
            -\mu_a & M & \mu_c
        \end{array}\right)
        \left(\begin{array}{ccc}
            j_b & J & j_d \\
            -\mu_b & -M & \mu_d
        \end{array}\right)~,
    \end{aligned}
\end{equation}
and the radial part
\begin{equation} \label{Eq:RadInt}
    \begin{aligned}
        R_J(ab;cd) = &(-1)^J(2J+1) \int dx_1~x_1^2 \int dx_2~x_2^2 \\
        &\left[\vphantom{\sum_L} C_J(\kappa_a,\kappa_c) C_J(\kappa_b,\kappa_d) W_{ac}(x_1)W_{bd}(x_2)g_J \right.\\
        &\left. + (-1)^{j_b+j_d}\sum_L X_{ca}(x_1)X_{bd}(x_2)g_L\right]~,
    \end{aligned}
\end{equation}
where
\begin{align}
    W_{ab}(x) =& f_a^1(x)f_b^1(x)+f_a^2(x)f_b^2(x)~, \\
    X_{ab}(x) =& f_a^1(x)f_b^2(x) S_{JL}(\kappa_a,-\kappa_b) \\
                 &- f_a^2(x)f_b^1(x) S_{JL}(-\kappa_a,\kappa_b)\nonumber~,
\end{align}
and $C_J(\kappa_a,\kappa_c)$ and $S_{JL}(\kappa_a,\kappa_b)$ are the reduced matrix elements as defined in Ref.~\cite{yas1999}.

\subsection{Subtraction Scheme}
\begin{figure}
    \centering
    \includegraphics[width=0.9\linewidth]{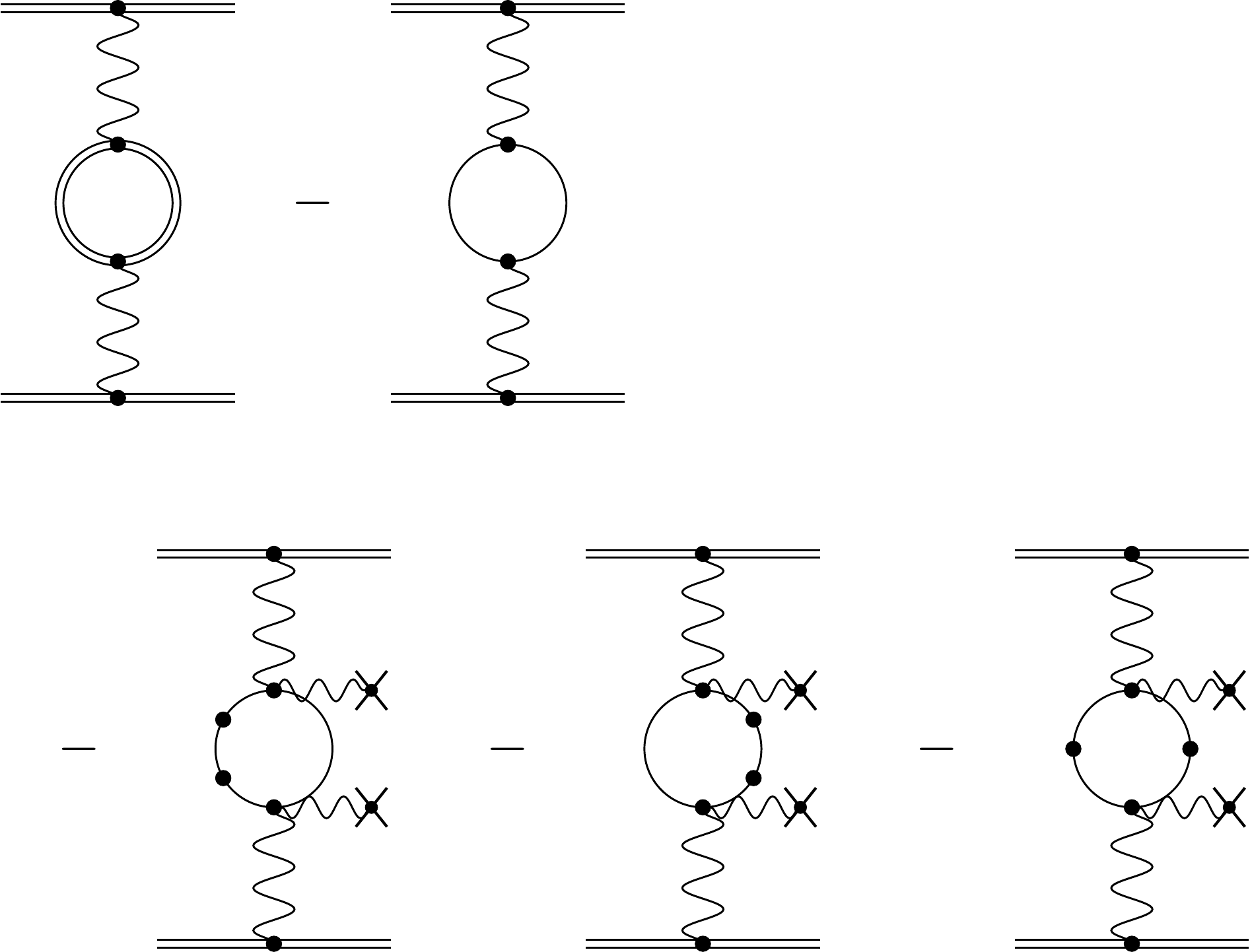}
    \caption{Subtraction scheme used to eliminate the lowest-order and spurious terms. The single line represents the free electron propagator, double lines denote the bound-electron propagator and wave function, wavy lines are photon propagators, wavy lines terminated by a cross denote the Coulomb potential, the single-dotted line corresponds to the propagator $-\frac{d}{dz}F_{ij}(z)$ and the double-dotted line to $\frac{1}{2}\frac{d^2}{dz^2}F_{ij}(z)$.}
    \label{Fig:FeynmanSub}
\end{figure}
It is convenient to split the energy shift~\eqref{Eq:EVP1} into the lowest-order Uehling term that has zero Coulomb interactions in the virtual electron-positron loop and the higher-order Wichmann-Kroll corrections with two or more interactions
\begin{equation}
    E_\text{$e$VP$e$}=E_\text{$e$VP$e$}^{(0)} + E_\text{$e$VP$e$}^{(2+)}~.
\end{equation}
The Uehling term can be obtained by replacing the Dirac-Coulomb propagators in Eq.~\eqref{Eq:EVP} with free Dirac propagators 
\begin{equation} \label{Eq:RepFree}
\begin{aligned}
    G_{23}&\left(z-\frac{\omega}{2}\right) G_{32}\left(z+\frac{\omega}{2}\right) \\
    &\to F_{23}\left(z-\frac{\omega}{2}\right)F_{32}\left(z+\frac{\omega}{2}\right)~,
\end{aligned}
\end{equation}
where  $G_{ij}(z) =  G\left(\boldsymbol{x}_i, \boldsymbol{x}_j, z\right)$ and  $F_{ij}(z) =  G\left(\boldsymbol{x}_i, \boldsymbol{x}_j, z\right)\vert_{Z=0}$. The renormalized Uehling contribution is well-known to be
\begin{equation}
    M^{(0)}(ab, cd; \omega) = \langle ab\vert U^{(0)}\vert cd \rangle~,
\end{equation}
where
\begin{equation}
\begin{aligned}
    U^{(0)}(\omega, \boldsymbol{x}_1, \boldsymbol{x}_2) = \frac{2\alpha^2}{3\pi} \frac{\alpha_{1\mu}\alpha^{\mu}_2}{\vert \boldsymbol{x}_1 - \boldsymbol{x}_2\vert} \int_1^\infty dt~\left(1 + \frac{1}{2t^2}\right)\\
    \times \frac{\sqrt{t^2-1}}{t^2} e^{-\sqrt{(2t)^2-\omega^2}\vert \boldsymbol{x}_1 - \boldsymbol{x}_2\vert}~.
\end{aligned}
\end{equation}
The corresponding calculation is straightforward and does not present any difficulties. For more details and calculations of the Uehling contribution, see Refs.~\cite{asy1997,abps1999}.

After the subtraction of the lowest-order Uehling term for each electron/positron multipole contribution, the Feynman diagram in Fig.~\ref{Fig:Feynman} is finite and, as shown by Artemyev and co-workers~\cite{asy1997,abps1999}, all spurious non-gauge-invariant terms vanish when summing over a finite number of multipoles and carrying out all the integrals. However, such spurious terms are still present before integration and give rise to contributions to the energy shifts that decrease only slowly with respect to the energy in the loop integral and the radial coordinates. Therefore, the convergence of the numerical integrals can be significantly accelerated by subtracting these terms explicitly. As was discussed by Scherdin and co-workers~\cite{ssgs1995} for the closely related process of Delbrück scattering, the spurious terms can be obtained from Eq.~\eqref{Eq:EVP} by replacing
\begin{equation} \label{Eq:RepRen}
\begin{aligned}
    G_{23}&\left(z-\frac{\omega}{2}\right) G_{32}\left(z+\frac{\omega}{2}\right) \\
    &\to \frac{1}{2}\frac{d^2}{dz^2}\left[ V_2F_{23}\left(z-\frac{\omega}{2}\right)F_{32}\left(z+\frac{\omega}{2}\right)V_3\right]~,
\end{aligned}
\end{equation}
where $V_i = V(x_i)$ is the Coulomb potential. Eq.~\eqref{Eq:RepRen} can be derived by replacing the electron mass with a large mass $M$ and taking the limit $M\to\infty$, see Refs.~\cite{gyu1975, sam1988}. Taking the derivative in Eq.~\eqref{Eq:RepRen} yields the three terms in the second line of Fig.~\ref{Fig:FeynmanSub}, where the single-dotted line corresponds to the propagator $-\frac{d}{dz}F_{ij}(z)$ and the double dotted line to $\frac{1}{2}\frac{d^2}{dz^2}F_{ij}(z)$.

Our numerical calculations confirm that the sum of the three spurious diagrams in Fig.~\ref{Fig:FeynmanSub} vanishes after integration for all individual multipole contributions, meaning that, in contrast to the self-energy~\cite{yhk2025, yhk2025a}, this subtraction cannot be used to accelerate the convergence of the multipole expansion. However, the numerical integrations over the radial coordinates and the loop-energy require significantly fewer integration points to achieve the same level of precision after subtracting the spurious terms. In our calculations, the accuracy of the final energy shift for a typical density of integration points increases by around two orders of magnitude due to this subtraction. 

\begin{figure}
    \centering
    \includegraphics[width=0.95\linewidth]{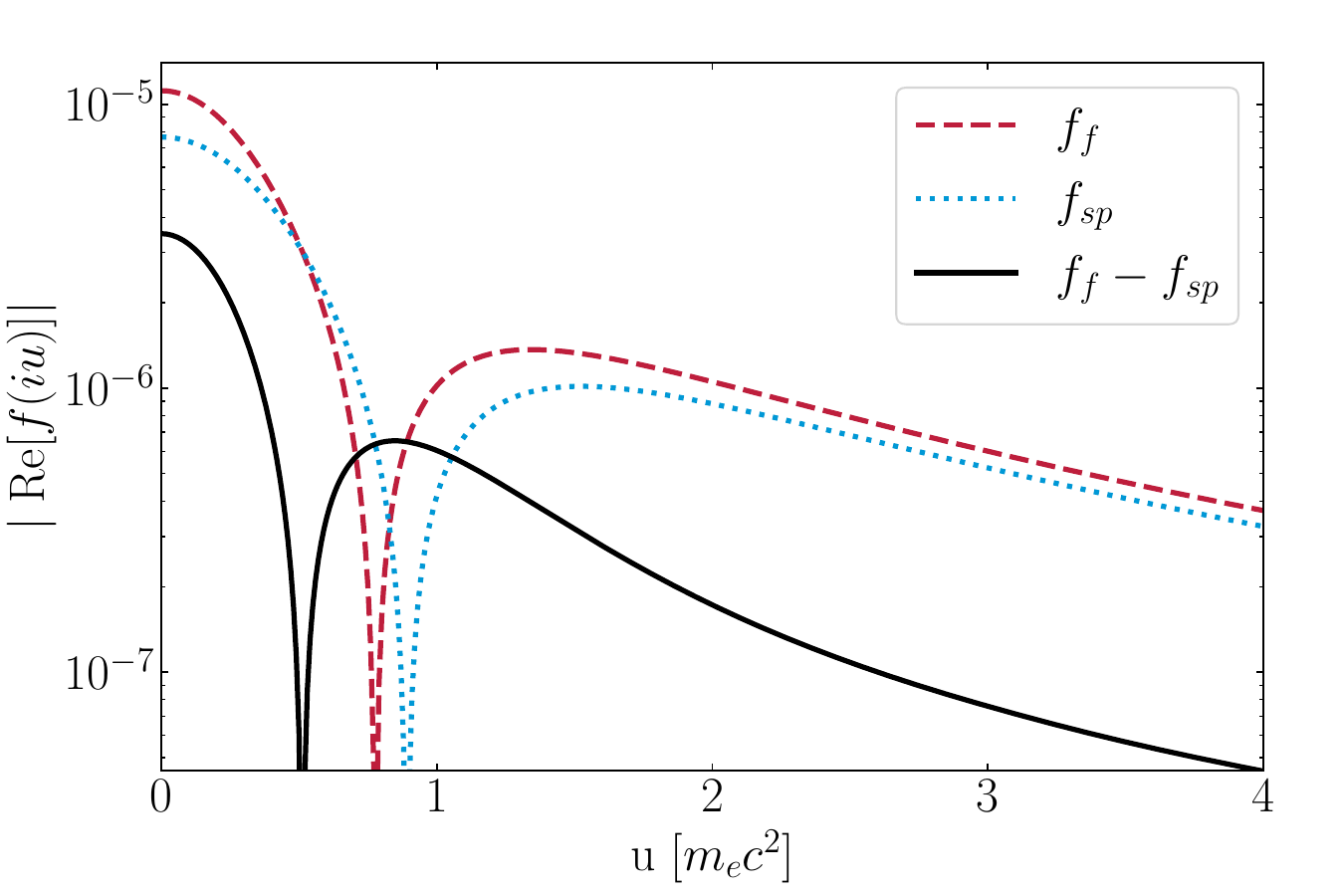}
    \caption{Absolute value of the real part of the integrand of the loop-energy integral along the imaginary axis for the ground state of He-like uranium including only the first term in the multipole expansion $\kappa$, $\kappa'$ = $\pm 1$. We display the full integrand $f_f$ that includes the spurious terms (red dashed line), the spurious terms $f_{sp}$ (blue dotted line), and the integrand without the spurious terms $f_f-f_{sp}$ (black solid line) for a point-like nucleus. The integral over $f_{sp}$ vanishes when carried out to infinity.}
    \label{Fig:Int}
\end{figure}

To visualize the effect of the spurious terms on the numerical convergence of the integrals, it is beneficial to look at the integrand $f$ of the loop-energy integral in Eq.~\eqref{Eq:EVP}, which we define by $E^{(2+)}_\text{$e$VP$e$} \equiv \int_{-\infty}^{+\infty}du~f(iu)$. In Fig.~\ref{Fig:Int}, we display this integrand for the ground state of He-like uranium for the first term in the multipole expansion of the virtual electron-positron pair. In the figure, $f_f$ corresponds to the full integrand, that includes the spurious terms and is described by the difference of the two Feynman diagrams in the first line of Fig.~\ref{Fig:FeynmanSub}, $f_{sp}$ is the spurious contribution described by the three Feynman diagrams in the second line of Fig.~\ref{Fig:FeynmanSub}, for which $\int_{-\infty}^{+\infty}du~f_{sp}(iu) = 0$, and $f_f - f_{sp}$ is the expression that is used in our calculations, where the spurious terms have been subtracted from the full integrand. As seen from the figure, the spurious terms contribute a non-physical slowly decreasing tail to the integrand, which worsens the numerical convergence. 

\section{Computational Details}~\label{Sec:Comp}
In the final expression for the interelectronic vacuum polarization correction in Eqs.~\eqref{Eq:AmpFinal}-\eqref{Eq:RadInt}, the integrals over the four radial coordinates and the energy of the virtual $e^+e^-$ loop remain to be evaluated numerically. After subtracting the spurious terms, these integrals converge relatively well. For the radial integrals, we use Gauß-Legendre quadrature to integrate the region close to the origin and then integrate the remaining part to infinity with Gauß-Laguerre integration. The Whittaker functions appearing in the integrand are evaluated using the FLINT library~\cite{flint, joh2017}. The integral over the loop-energy $z$ in Eq.~\eqref{Eq:AmpFinal} is performed using a Wick rotation to integrate along the imaginary axis and again using Gauß-Legendre quadrature. We choose the number and density of the integration points large enough so that increasing the number of points by 50\% results in a relative change in the final energy shift of less than $10^{-6}$.

The largest numerical uncertainty originates from the multipole expansion of the two electron propagators, which contains infinitely many terms and has to be truncated at some point. In our calculations, we sum over all terms up to $\kappa, \kappa'= \pm 16$ and estimate the remaining tail by extrapolating. This extrapolation is performed separately along the diagonals where $\vert\kappa\vert = \vert\kappa'\vert$, $\vert\kappa\vert = \vert\kappa'\vert \pm 1$, $\vert\kappa\vert = \vert\kappa'\vert \pm 2$, ... with only a finite number of these diagonals being non-zero due to the angular momentum selection rules for the transition between the bound electronic states. The remaining tail is estimated by fitting polynomials in inverse powers of $\vert\kappa\vert$ of different degree to the terms of the expansion, where the uncertainty of this procedure is obtained by comparing how the result changes when the truncation point is varied by 25\% and when the number of terms in the polynomial that is fitted increases. This uncertainty is explicitly given for the results presented in the next section. 

For the finite nuclear size correction, we calculate the difference in the energy shift between an extended nucleus and the point-like case $\Delta E^{(2+)}_\text{$e$VP$e$}(\text{Fns}) = E^{(2+)}_\text{$e$VP$e$}(\text{Ext})-E^{(2+)}_\text{$e$VP$e$}(\text{Pnt})$ for each multipole contribution and sum these differences. As expected, the convergence of $\Delta E^{(2+)}_\text{$e$VP$e$}(\text{Fns})$ with respect to $\kappa, \kappa'$ is significantly faster and no extrapolation is necessary to achieve sufficient numerical precision. This is because electronic states with large angular momenta have little overlap with the nucleus, meaning that the Coulomb potential seen by these states is very close to that created by a point-like charge distribution. We compare the results obtained when employing a spherical shell, homogeneously charged sphere, and Fermi charge distribution to estimate the nuclear model dependence, which is included in the theoretical uncertainty of the finite nuclear size shift. For each charge number $Z$, the experimentally determined nuclear charge radius and its uncertainty are taken from Ref.~\cite{aam2013} for the most abundant isotope if available, and otherwise calculated from the standard approximation formula $R=(0.836A^{1/3}+0.570)~\si{\femto \meter}$~\cite{jas1985}.

\section{Numerical Results}~\label{Sec:Res}
\begin{table}
\begin{ruledtabular}
\begin{tabular}{cddd}
\\[-8pt]
& \multicolumn{3}{c}{$(1s_{1/2})^2$}\\
$Z$ & \multicolumn{1}{c}{Pnt} & \multicolumn{1}{c}{Fns} & \multicolumn{1}{c}{Ref.~\cite{asy1997}}\\
\\[-8pt]
\colrule
\\[-6pt]
$20$ & -0.00051(1) & 0.00000(1) & -0.\\
\\[-6pt]
$30$ & -0.00602(2) & 0.00000(1) & -0.\\
\\[-6pt]
$40$ & -0.0303(1) & 0.0000(1) & -0.\\
\\[-6pt]
$50$ & -0.0908(1) & -0.0002(1) & -0.\\
\\[-6pt]
$54$ & -0.1249(1) & -0.0004(1) & -0.\\
\\[-6pt]
$60$ & -0.176(1) & -0.001(1) & -0.\\
\\[-8pt]
$74$ & -0.047(1) & -0.013(1) & -0.\\
\\[-8pt]
$80$ & 0.366(1) & -0.032(2) & 0.\\
\\[-8pt]
$83$ & 0.748(2) & -0.049(2) & 1.\\
\\[-8pt]
$90$ & 2.387(3) & -0.138(3) & 2.\\
\\[-8pt]
$92$ & 3.126(3) & -0.184(3) & 3.\\
\\[-8pt]
$100$ & 7.948(8) & -0.528(4) & 8.\\
\end{tabular}
\end{ruledtabular}
\caption{Wichmann-Kroll correction to the interelectronic interaction in Fig.~\ref{Fig:Feynman} for the ground state of helium-like ions in meV. We present the correction for a point-like nucleus (Pnt) and the additional shift due to finite nuclear size effects (Fns). Theoretical uncertainties are given in parentheses.} \label{Tab:He}
\end{table}

\begin{table*}
\begin{ruledtabular}
\begin{tabular}{cdddddd}
\\[-8pt]
& \multicolumn{2}{c}{$(1s_{1/2})^22s_{1/2}$} & \multicolumn{2}{c}{$(1s_{1/2})^22p_{1/2}$} & \multicolumn{2}{c}{$(1s_{1/2})^22p_{3/2}$}\\
$Z$ & \multicolumn{1}{c}{Pnt} & \multicolumn{1}{c}{Fns} & \multicolumn{1}{c}{Pnt} & \multicolumn{1}{c}{Fns} & \multicolumn{1}{c}{Pnt} & \multicolumn{1}{c}{Fns} \\
\\[-8pt]
\colrule
\\[-6pt]
$20$ & -0.00006(1) & 0.00000(1) & 0.00004(1) & 0.00000(1) & 0.00005(1) & 0.00000(1)\\
\\[-6pt]
$30$ & -0.00070(1) & 0.00000(1) & 0.00027(1) & 0.00000(1) & 0.00034(1) & 0.00000(1)\\
\\[-6pt]
$40$ & -0.0035(1) & 0.0000(1) & 0.00029(1) & 0.00003(1) & 0.0012(1) & 0.0000(1)\\
\\[-8pt]
$50$ & -0.0100(1) & 0.0000(1) & -0.0045(1) & 0.0001(1) & 0.0020(1) & 0.0000(1)\\
\\[-8pt]
$54$ & -0.0135(1) & -0.0001(1) & -0.0114(1) & 0.0002(1) & 0.0015(1) & 0.0000(1)\\
\\[-6pt]
$60$ & -0.0180(1) & -0.0002(1) & -0.0347(1) & 0.0004(1) & -0.0017(1) & 0.0000(1)\\
\\[-8pt]
$74$ & 0.0056(1) & -0.002(1) & -0.274(2) & 0.002(1) & -0.0439(1) & 0.0002(1)\\
\\[-8pt]
$80$ & 0.0600(1) & -0.005(1) & -0.580(1) & 0.005(2) & -0.0943(1) & 0.0004(2)\\
\\[-8pt]
$83$ & 0.1071(1) & -0.007(1) & -0.833(1) & 0.007(3) & -0.134(1) & 0.0005(2)\\
\\[-8pt]
$90$ & 0.294(1) & -0.020(2) & -1.878(1) & 0.022(4) & -0.284(2) & 0.002(1)\\
\\[-8pt]
$92$ & 0.373(1) & -0.027(2) & -2.356(1) & 0.030(6) & -0.347(1) & 0.002(1)\\
\\[-8pt]
$92$\footnote{Ref.~\cite{abps1999}} & -2.1 & - & -1.3 & - & -0.2 & -\\
\\[-8pt]
$100$ & 0.820(1) & -0.073(2) & -5.758(1) & 0.115(10) & -0.738(1) & 0.007(2)
\end{tabular}
\end{ruledtabular}
\caption{Wichmann-Kroll correction to the interelectronic interaction in Fig.~\ref{Fig:Feynman} due to the interaction of the valence electron in the $2s_{1/2}$, $2p_{1/2}$ or $2p_{3/2}$ state of a Li-like ion with the helium-like core in meV. We present the correction for a point-like nucleus (Pnt) and the additional shift due to finite nuclear size effects (Fns). Theoretical uncertainties are given in parentheses.} \label{Tab:Li}
\end{table*}

We are now ready to calculate the contribution of the Feynman diagram in Fig.~\eqref{Fig:Feynman} to the Lamb shift in highly charged ions. We start our discussion with the ground state of two-electron ions. The numerical results of our calculation for different nuclear charge numbers $Z$ along with rounded up theoretical uncertainties from the numerical procedures and nuclear structure are presented in Table~\ref{Tab:He}. As shown in the table, the shift is relatively small and amounts to only a few \si{\milli\eV}, even for heavy ions. We observe a $Z^6$ scaling of the energy shift in the low-$Z$ region. Moreover, it is interesting to note that the correction changes sign near $Z = 74$. Our results are in excellent agreement with the previous calculations of Artemyev and co-workers~\cite{asy1997} and improve the accuracy by several orders of magnitude.

The finite nuclear size correction generally reduces the size of the vacuum polarization interelectronic correction since the Coulomb potential inside an extended nucleus is reduced. Although the overall size of $\Delta E^{(2+)}_\text{$e$VP$e$}(\text{Fns})$ is again small and below $1~\si{\milli\eV}$, its relative size compared to the point-like case can exceed 5\% for high-$Z$ ions. The results depend only weakly on the choice of the nuclear model, with even the relatively simple spherical shell model yielding accurate results. For example, for helium-like uranium, the relative difference between $E^{(2+)}_\text{$e$VP$e$}(\text{Ext})$ evaluated for the spherical shell and the Fermi model is less than 0.1\%.

We now consider the shift in lithium-like highly charged ions. The additional shift arising from the interaction of the valence electron occupying either the $2s_{1/2}$, $2p_{1/2}$, or $2p_{3/2}$ state, with the two electrons in the $1s_{1/2}$ state is presented in Table~\ref{Tab:Li}. In this case, it is necessary to sum over both angular momentum projections of the $1s_{1/2}$ electrons, since both states are occupied
\begin{equation}
    E_\text{$e$VP$e$} = \sum_{\mu_a} M(ab,ab;0) - M(ba,ab;\omega)~.
\end{equation}
As shown in the table, the shift is similar in size to the helium-like case, amounting to a few \si{\milli\eV}. Moreover, as before, the shift changes sign when going from low- to high-$Z$ ions. While for the $2s_{1/2}$ state, the sign again changes from negative to positive, the opposite is true for the $2p$ states. For lithium-like uranium, we compare our results with the calculation of Artemyev and co-workers~\cite{abps1999}, for which we find a discrepancy between the results. Despite extensive investigation, including numerical verification of all steps in the derivation to arrive at the final analytical expressions, we were unable to identify the origin of this discrepancy. However, our results are in very good agreement with independent calculations that were provided by A. Malyshev~\cite{mal2023}. The finite nuclear size correction again generally reduces the size of the higher-order two-electron correction by more than 5\% in the high-$Z$ regime.

\section{Conclusions}~\label{Sec:Conc}
We have presented a theoretical study of the higher-order vacuum polarization correction to the interelectronic interaction in highly charged He- and Li-like ions. We have discussed the basic formalism for the computation in the framework of bound-state QED, placing a special emphasis on the presence of non-gauge-invariant spurious terms that appear in the calculation. Although these terms vanish after all integrations are performed, they are still present in the integrands and slow down numerical convergence. We have implemented a method to subtract the spurious terms explicitly to obtain expressions that can be numerically evaluated more efficiently. We have performed extensive calculations of the Wichmann-Kroll correction to the interelectronic interaction for a wide range of nuclear charge numbers $Z$. Our calculations are in very good agreement with previous results by Artemyev and co-workers for the case of He-like ions, but we observe a discrepancy for the Li-like case. The calculated correction is on the order of a few meV and has a positive sign for the $s$ states in the high-$Z$ regime and a negative sign for low-$Z$ ions, while the inverse is the case for the $p$ states. Moreover, we have explicitly discussed the influence of the finite nuclear size, which can reduce the effect by more than 5\% for heavy HCIs. The methods developed in this work can be adapted to the more complicated two-loop QED diagram, where a virtual electron-positron loop is inserted into the self-energy photon line. An all-order calculation of this diagram is of great interest as it currently limits the accuracy of theoretical predictions in highly-charged ions~\cite{yas2015}.

\section*{Acknowledgements} 
This work was supported by the German Research Foundation (Deutsche Forschungsgemeinschaft, DFG) under project SO 2403/2-1. We thank A. Malyshev for valuable discussions and for the data comparison.

\appendix

\section{Interelectronic Interaction in the Redefined Vacuum} \label{Ap:RedVac}
The redefined vacuum approach can simplify the treatment of interactions involving core electrons by defining the Fermi energy $E_F$ to be above the energy of the highest core state. Considering an HCI with two electrons in states $a$ and $b$ beyond the closed shell $E_{a,b}>E_F$, we can recover the interactions with the core electrons using the following identity for the difference of the electron propagator in the redefined and standard vacuum
\begin{equation} \label{Eq:RVCut}
\begin{aligned}
\sum_{n\kappa\mu} &\frac{\psi_{n\kappa\mu}(\boldsymbol{x}_1)\psi^\dagger_{n\kappa\mu}(\boldsymbol{x}_2)}{E_{n\kappa}-z-i\delta(E_{n\kappa}-E_F)} - \sum_{n\kappa\mu} \frac{\psi_{n\kappa\mu}(\boldsymbol{x}_1)\psi^\dagger_{n\kappa\mu}(\boldsymbol{x}_2)}{E_{n\kappa}-z-i\delta} \\
&= 2\pi i \sum_{E_{n\kappa} < E_F} \delta(z-E_{n\kappa}) \psi_{n\kappa\mu}(\boldsymbol{x}_1)\psi^\dagger_{n\kappa\mu}(\boldsymbol{x}_2)~,
\end{aligned}
\end{equation}
which is to be understood under the integral over $z$. The replacement of an electron propagator with this difference corresponds to cutting the respective electron line in the Feynman diagram, see Refs.~\cite{sha2002,svtg2021} for more details.

Applying Eq.~\eqref{Eq:RVCut} to the exchange term $M(ba,ab;\omega)$ in Eq.~\eqref{Eq:EVP1} yields two contributions, depending on which propagator in Eq.~\eqref{Eq:EVP} is replaced:
\begin{widetext}
\begin{equation} \label{Eq:RV1}
\begin{aligned}
   M_L(ba,ab;\omega) = (4\pi\alpha)^2 \sum_{E_{n\kappa} < E_F}\int &d^3\boldsymbol{x}_1~d^3\boldsymbol{x}_2~ d^3\boldsymbol{x}_3~d^3\boldsymbol{x}_4~\psi^\dagger_b(\boldsymbol{x}_1)\alpha_\mu\psi_a(\boldsymbol{x}_1) D^{\mu\nu}(\omega,\boldsymbol{x}_{12})\\
    &\times\psi^\dagger_{n\kappa\mu}(\boldsymbol{x}_3) \alpha_\rho G\left(\boldsymbol{x}_3, \boldsymbol{x}_2, E_{n\kappa}+\omega\right)\alpha_\nu \psi_{n\kappa\mu}(\boldsymbol{x}_2) D^{\rho\sigma}(\omega,\boldsymbol{x}_{34}) \psi^\dagger_a(\boldsymbol{x}_4)\alpha_\sigma\psi_b(\boldsymbol{x}_4)~,
\end{aligned}
\end{equation}
and
\begin{equation} \label{Eq:RV2}
\begin{aligned}
   M_R(ba,ab;\omega) = (4\pi\alpha)^2 \sum_{E_{n\kappa} < E_F}\int &d^3\boldsymbol{x}_1~d^3\boldsymbol{x}_2~ d^3\boldsymbol{x}_3~d^3\boldsymbol{x}_4~\psi^\dagger_b(\boldsymbol{x}_1)\alpha_\mu\psi_a(\boldsymbol{x}_1) D^{\mu\nu}(\omega,\boldsymbol{x}_{12})\\
    &\times\psi^\dagger_{n\kappa\mu}(\boldsymbol{x}_2)\alpha_\nu G\left(\boldsymbol{x}_2, \boldsymbol{x}_3, E_{n\kappa}-\omega\right) \alpha_\rho \psi_{n\kappa\mu}(\boldsymbol{x}_3) D^{\rho\sigma}(\omega,\boldsymbol{x}_{34}) \psi^\dagger_a(\boldsymbol{x}_4)\alpha_\sigma\psi_b(\boldsymbol{x}_4)~.
\end{aligned}
\end{equation}
\end{widetext}

\begin{figure}
\center
\includegraphics[scale=0.45]{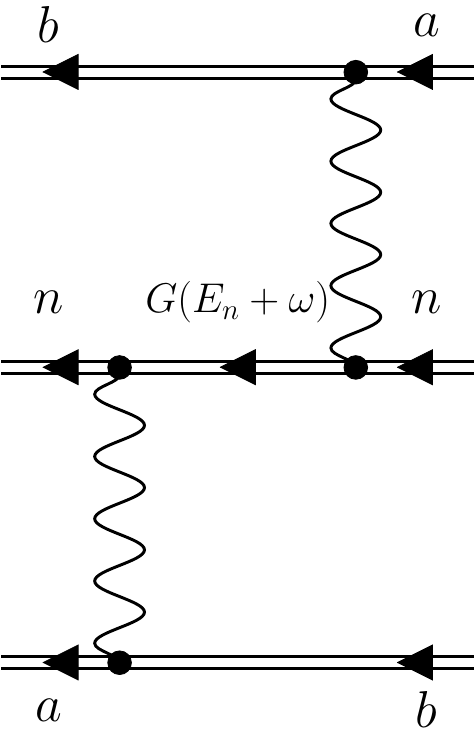}
\hspace{0.5cm}
\includegraphics[scale=0.45]{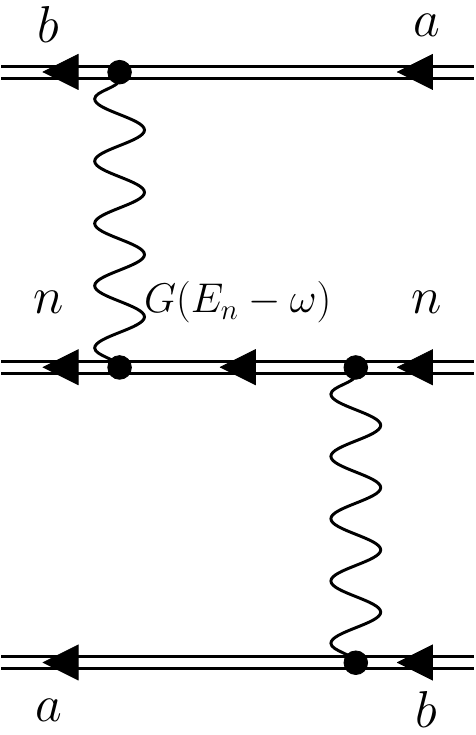}
\caption{Three-electron two-photon exchange Feynman diagrams corresponding to Eq.~\eqref{Eq:RV1} (left) and Eq.~\eqref{Eq:RV2} (right).} \label{Fig:FeynmanRV}
\end{figure}

These two contributions correspond to the three-electron two-photon exchange diagrams shown in Fig.~\ref{Fig:FeynmanRV}. From this connection, it is obvious that with the sign convention in the energy argument of the electron Green's functions as used in Eqs.~\eqref{Eq:EVP}, \eqref{Eq:RV1} and \eqref{Eq:RV2}, only $\omega = E_a-E_b$ gives a result consistent with the energy conservation requirements at each vertex. This fact was pointed out to us by Aleksei Malyshev~\cite{mal2023}.

%%%%%%%%%%%%%%%%%%%%%%%%%%%%%%%%%%%%%%%%%%%%%%%%%%%%%%%%%%%%%%%%%%%%%%%%%%%%%%%%%%%%%%%%%%%

\bibliography{./eVPe.references}

@article{asy1997,
  title = {{Vacuum polarization screening corrections to the ground-state energy of two-electron ions}},
  author = {Artemyev, A. N. and Shabaev, V. M. and Yerokhin, V. A.},
  journal = {Phys. Rev. A},
  volume = {56},
  issue = {5},
  pages = {3529--3534},
  numpages = {0},
  year = {1997},
  month = {Nov},
  publisher = {American Physical Society},
  doi = {10.1103/PhysRevA.56.3529},
  url = {https://link.aps.org/doi/10.1103/PhysRevA.56.3529}
}

@article{ind2019,
   Author = {P Indelicato},
   Title = {\relax{Topical Review: QED tests with highly-charged ions}},
   Journal = {jpbAMOP},
   Volume = {52},
   Pages = {232001},
      URL = {https://doi.org/10.1088/1361-6455/ab42c9},
   Year = {2019} }

@article{lbds2024,
   Author = {R. Loetzsch and H. F. Beyer and L. Duval and U. Spillmann and D. Banaś and P. Dergham and F. M. Kröger and J. Glorius and R. E. Grisenti and M. Guerra and A. Gumberidze and R. Heß and P. M. Hillenbrand and P. Indelicato and P. Jagodzinski and E. Lamour and B. Lorentz and S. Litvinov and Yu A. Litvinov and J. Machado and N. Paul and G. G. Paulus and N. Petridis and J. P. Santos and M. Scheidel and R. S. Sidhu and M. Steck and S. Steydli and K. Szary and S. Trotsenko and I. Uschmann and G. Weber and Th Stöhlker and M. Trassinelli},
   Title = {{Testing Quantum Electrodynamics in Extreme Fields Using Helium-Like Uranium}},
   Journal = {Nature},
   Volume = {625},
   Number = {7996},
   Pages = {673-678},
      URL = {https://doi.org/10.1038/s41586-023-06910-y
https://www.nature.com/articles/s41586-023-06910-y.pdf},
   Year = {2024} }

@Article{sdhh2022,
author={Sailer, Tim
and Debierre, Vincent
and Harman, Zolt{\'a}n
and Hei{\ss}e, Fabian
and K{\"o}nig, Charlotte
and Morgner, Jonathan
and Tu, Bingsheng
and Volotka, Andrey V.
and Keitel, Christoph H.
and Blaum, Klaus
and Sturm, Sven},
title={{Measurement of the bound-electron g-factor difference in coupled ions}},
journal={Nature},
year={2022},
month={Jun},
day={01},
volume={606},
number={7914},
pages={479-483},
issn={1476-4687},
doi={10.1038/s41586-022-04807-w},
url={https://doi.org/10.1038/s41586-022-04807-w}
}

@Article{sbgk2018,
author={Shabaev, V. M.
and Bondarev, A. I.
and Glazov, D. A.
and Kaygorodov, M. Y.
and Kozhedub, Y. S.
and Maltsev, I. A.
and Malyshev, A. V.
and Popov, R. V.
and Tupitsyn, I. I.
and Zubova, N. A.},
title={{Stringent tests of QED using highly charged ions}},
journal={Hyperfine Interactions},
year={2018},
month={Nov},
day={26},
volume={239},
number={1},
pages={60},
issn={1572-9540},
doi={10.1007/s10751-018-1537-8},
url={https://doi.org/10.1007/s10751-018-1537-8}
}

@Article{uabd2017,
author={Ullmann, Johannes
and Andelkovic, Zoran
and Brandau, Carsten
and Dax, Andreas
and Geithner, Wolfgang
and Geppert, Christopher
and Gorges, Christian
and Hammen, Michael
and Hannen, Volker
and Kaufmann, Simon
and K{\"o}nig, Kristian
and Litvinov, Yuri A.
and Lochmann, Matthias
and Maa{\ss}, Bernhard
and Meisner, Johann
and Murb{\"o}ck, Tobias
and S{\'a}nchez, Rodolfo
and Schmidt, Matthias
and Schmidt, Stefan
and Steck, Markus
and St{\"o}hlker, Thomas
and Thompson, Richard C.
and Trageser, Christian
and Vollbrecht, Jonas
and Weinheimer, Christian
and N{\"o}rtersh{\"a}user, Wilfried},
title={{High precision hyperfine measurements in Bismuth challenge bound-state strong-field QED}},
journal={Nature Communications},
year={2017},
month={May},
day={16},
volume={8},
number={1},
pages={15484},
issn={2041-1723},
doi={10.1038/ncomms15484},
url={https://doi.org/10.1038/ncomms15484}
}

@article{gsbb2004,
  title = {{Electron-Electron Interaction in Strong Electromagnetic Fields: The Two-Electron Contribution to the Ground-State Energy in He-like Uranium}},
  author = {Gumberidze, A. and St\"ohlker, Th. and Bana\ifmmode \acute{s}\else \'{s}\fi{}, D. and Beckert, K. and Beller, P. and Beyer, H. F. and Bosch, F. and Cai, X. and Hagmann, S. and Kozhuharov, C. and Liesen, D. and Nolden, F. and Ma, X. and Mokler, P. H. and Or\ifmmode\check{s}\else\v{s}\fi{}i\ifmmode\acute{c}\else\'{c}\fi{}-Muthig, A. and Steck, M. and Sierpowski, D. and Tashenov, S. and Warczak, A. and Zou, Y.},
  journal = {Phys. Rev. Lett.},
  volume = {92},
  issue = {20},
  pages = {203004},
  numpages = {4},
  year = {2004},
  month = {May},
  publisher = {American Physical Society},
  doi = {10.1103/PhysRevLett.92.203004},
  url = {https://link.aps.org/doi/10.1103/PhysRevLett.92.203004}
}

@article{asyp2005,
  title = {{QED calculation of the $n=1$ and $n=2$ energy levels in He-like ions}},
  author = {Artemyev, A. N. and Shabaev, V. M. and Yerokhin, V. A. and Plunien, G. and Soff, G.},
  journal = {Phys. Rev. A},
  volume = {71},
  issue = {6},
  pages = {062104},
  numpages = {26},
  year = {2005},
  month = {Jun},
  publisher = {American Physical Society},
  doi = {10.1103/PhysRevA.71.062104},
  url = {https://link.aps.org/doi/10.1103/PhysRevA.71.062104}
}

@article{pcjy2005,
  title = {{Complete two-loop correction to the bound-electron $g$ factor}},
  author = {Pachucki, Krzysztof and Czarnecki, Andrzej and Jentschura, Ulrich D. and Yerokhin, Vladimir A.},
  journal = {Phys. Rev. A},
  volume = {72},
  issue = {2},
  pages = {022108},
  numpages = {11},
  year = {2005},
  month = {Aug},
  publisher = {American Physical Society},
  doi = {10.1103/PhysRevA.72.022108},
  url = {https://link.aps.org/doi/10.1103/PhysRevA.72.022108}
}

@article{sykh2025,
  title = {{Improved Bound-Electron $g$-Factor Theory through Complete Two-Loop QED Calculations}},
  author = {Sikora, B. and Yerokhin, V. A. and Keitel, C. H. and Harman, Z.},
  journal = {Phys. Rev. Lett.},
  volume = {134},
  issue = {12},
  pages = {123001},
  numpages = {6},
  year = {2025},
  month = {Mar},
  publisher = {American Physical Society},
  doi = {10.1103/PhysRevLett.134.123001},
  url = {https://link.aps.org/doi/10.1103/PhysRevLett.134.123001}
}

@article{yas2015,
    author = {Yerokhin, V. A. and Shabaev, V. M.},
    title = {{Lamb Shift of n = 1 and n = 2 States of Hydrogen-like Atoms, 1 $\leq$ Z $\leq$ 110}},
    journal = {J. Phys. Chem. Ref. Data.},
    volume = {44},
    number = {3},
    pages = {033103},
    year = {2015},
    month = {08},
    issn = {0047-2689},
    doi = {10.1063/1.4927487},
    url = {https://doi.org/10.1063/1.4927487},
}

@article{raw1973,
  title = {{Vacuum Polarization in High-$Z$, Finite-Size Nuclei}},
  author = {Rinker, G. A. and Wilets, L.},
  journal = {Phys. Rev. Lett.},
  volume = {31},
  issue = {26},
  pages = {1559--1562},
  numpages = {0},
  year = {1973},
  month = {Dec},
  publisher = {American Physical Society},
  doi = {10.1103/PhysRevLett.31.1559},
  url = {https://link.aps.org/doi/10.1103/PhysRevLett.31.1559}
}

@article{gyu1974,
  title = {{Nuclear-Size Effects on Vacuum Polarization in Muonic Pb}},
  author = {Gyulassy, M.},
  journal = {Phys. Rev. Lett.},
  volume = {32},
  issue = {24},
  pages = {1393--1396},
  numpages = {0},
  year = {1974},
  month = {Jun},
  publisher = {American Physical Society},
  doi = {10.1103/PhysRevLett.32.1393},
  url = {https://link.aps.org/doi/10.1103/PhysRevLett.32.1393}
}

@article{sam1988,
   Author = {G. Soff and P.J. Mohr},
   Title = {\relax{Vacuum Polarization in a Strong External Field}},
   Journal = pra,
   Volume = {38},
   Number = {10},
   Pages = {5066},
    Year = {1988}
}

@article{moh1974,
title = {{Self-energy radiative corrections in hydrogen-like systems}},
journal = {Ann. Phys.},
volume = {88},
number = {1},
pages = {26-51},
year = {1974},
issn = {0003-4916},
doi = {https://doi.org/10.1016/0003-4916(74)90398-4},
url = {https://www.sciencedirect.com/science/article/pii/0003491674903984},
author = {Peter J Mohr}
}

@article{moh1982,
  title = {{Self-energy of the $n=2$ states in a strong Coulomb field}},
  author = {Mohr, Peter J.},
  journal = {Phys. Rev. A},
  volume = {26},
  issue = {5},
  pages = {2338--2354},
  numpages = {0},
  year = {1982},
  month = {Nov},
  publisher = {American Physical Society},
  doi = {10.1103/PhysRevA.26.2338},
  url = {https://link.aps.org/doi/10.1103/PhysRevA.26.2338}
}

@article{yis2008,
  title = {{Two-loop QED corrections with closed fermion loops}},
  author = {Yerokhin, V. A. and Indelicato, P. and Shabaev, V. M.},
  journal = {Phys. Rev. A},
  volume = {77},
  issue = {6},
  pages = {062510},
  numpages = {12},
  year = {2008},
  month = {Jun},
  publisher = {American Physical Society},
  doi = {10.1103/PhysRevA.77.062510},
  url = {https://link.aps.org/doi/10.1103/PhysRevA.77.062510}
}

@article{mks2023,
  title = {{Ab initio calculations of the $2{p}_{3/2}\ensuremath{\rightarrow}2s$ transition in He-, Li-, and Be-like uranium}},
  author = {Malyshev, A. V. and Kozhedub, Y. S. and Shabaev, V. M.},
  journal = {Phys. Rev. A},
  volume = {107},
  issue = {4},
  pages = {042806},
  numpages = {12},
  year = {2023},
  month = {Apr},
  publisher = {American Physical Society},
  doi = {10.1103/PhysRevA.107.042806},
  url = {https://link.aps.org/doi/10.1103/PhysRevA.107.042806}
}

@article{syss2023,
  title = {{All-Order Coulomb Corrections to Delbr\"uck Scattering above the Pair-Production Threshold}},
  author = {Sommerfeldt, J. and Yerokhin, V. A. and St\"ohlker, Th. and Surzhykov, A.},
  journal = {Phys. Rev. Lett.},
  volume = {131},
  issue = {6},
  pages = {061601},
  numpages = {6},
  year = {2023},
  month = {Aug},
  publisher = {American Physical Society},
  doi = {10.1103/PhysRevLett.131.061601},
  url = {https://link.aps.org/doi/10.1103/PhysRevLett.131.061601}
}

@article{aylv2026,
  title = {{Energy spectra of electron-positron pairs produced in bound-bound muon transitions}},
  author = {Andreev, Oleg Yu. and Yu, Deyang and Lyashchenko, Konstantin N. and Vasileva, Daria M.},
  journal = {Phys. Rev. A},
  volume = {113},
  issue = {4},
  pages = {042804},
  numpages = {8},
  year = {2026},
  month = {Apr},
  publisher = {American Physical Society},
  doi = {10.1103/m9l8-6g96},
  url = {https://link.aps.org/doi/10.1103/m9l8-6g96}
}

@article{abps1999,
  title = {{Vacuum-polarization screening corrections to the energy levels of lithiumlike ions}},
  author = {Artemyev, A. N. and Beier, T. and Plunien, G. and Shabaev, V. M. and Soff, G. and Yerokhin, V. A.},
  journal = {Phys. Rev. A},
  volume = {60},
  issue = {1},
  pages = {45--49},
  numpages = {0},
  year = {1999},
  month = {Jul},
  publisher = {American Physical Society},
  doi = {10.1103/PhysRevA.60.45},
  url = {https://link.aps.org/doi/10.1103/PhysRevA.60.45}
}

@phdthesis{sun1998,
  title        = {{Complete One-Loop QED Calculations for Few-Electrons Ions. - Applications to Electron-Electron Interaction, the Zeeman Effect and Hyperfine Structure}},
  author       = {P. Sunnergren},
  year         = {1998},
  school       = {University of Gothenburg},
  type         = {PhD thesis}
}

@article{sha2002,
title = {{Two-time Green's function method in quantum electrodynamics of high-Z few-electron atoms}},
journal = {Phys. Rep.},
volume = {356},
number = {3},
pages = {119-228},
year = {2002},
issn = {0370-1573},
doi = {https://doi.org/10.1016/S0370-1573(01)00024-2},
url = {https://www.sciencedirect.com/science/article/pii/S0370157301000242},
author = {V.M. Shabaev}
}

@article{hyl1984,
    author = {Hylton, D. J.},
    title = {{The reduced Dirac Green function for the Coulomb potential}},
    journal = {J. Math. Phys.},
    volume = {25},
    number = {4},
    pages = {1125-1132},
    year = {1984},
    month = {04},
    issn = {0022-2488},
    doi = {10.1063/1.526255},
    url = {https://doi.org/10.1063/1.526255},
}

@article{mps1998,
   Author = {P.J. Mohr and G. Plunien and G. Soff},
   Title = {{QED Corrections in Heavy Atoms}},
   Journal = {Phys. Rep.},
   Volume = {293},
   Number = {5&6},
   Pages = {227-372},
      URL = {https://doi.org/10.1016/S0370-1573(97)00046-X},
   Year = {1998} 
}

@article{sap2026,
  title = {{All-order Wichmann-Kroll contribution in heavy electronic and exotic atoms}},
  author = {Sommerfeldt, Jonas and Indelicato, Paul},
  journal = {Phys. Rev. A},
  volume = {113},
  issue = {2},
  pages = {022806},
  numpages = {8},
  year = {2026},
  month = {Feb},
  publisher = {American Physical Society},
  doi = {10.1103/ylft-n4f9},
  url = {https://link.aps.org/doi/10.1103/ylft-n4f9}
}

@article{sam1991,
title = {{Accurate numerical solution of the Schrödinger and Dirac wave equations for central fields}},
journal = {Comput. Phys. Commun.},
volume = {62},
number = {1},
pages = {65-79},
year = {1991},
issn = {0010-4655},
doi = {https://doi.org/10.1016/0010-4655(91)90122-2},
url = {https://www.sciencedirect.com/science/article/pii/0010465591901222},
author = {Francesc Salvat and Ricardo Mayol}
}

@article{yas1999,
  title = {{First-order self-energy correction in hydrogenlike systems}},
  author = {Yerokhin, V. A. and Shabaev, V. M.},
  journal = {Phys. Rev. A},
  volume = {60},
  issue = {2},
  pages = {800--811},
  numpages = {0},
  year = {1999},
  month = {Aug},
  publisher = {American Physical Society},
  doi = {10.1103/PhysRevA.60.800},
  url = {https://link.aps.org/doi/10.1103/PhysRevA.60.800}
}

@article{jbs1988,
  title = {{Many-body perturbation-theory calculations of energy levels along the lithium isoelectronic sequence}},
  author = {Johnson, W. R. and Blundell, S. A. and Sapirstein, J.},
  journal = {Phys. Rev. A},
  volume = {37},
  issue = {8},
  pages = {2764--2777},
  numpages = {0},
  year = {1988},
  month = {Apr},
  publisher = {American Physical Society},
  doi = {10.1103/PhysRevA.37.2764},
  url = {https://link.aps.org/doi/10.1103/PhysRevA.37.2764}
}

@Article{ssgs1995,
author={Scherdin, A.
and Sch{\"a}fer, A.
and Greiner, W.
and Soff, G.
and Mohr, P. J.},
title={{Coulomb corrections to Delbr{\"u}ck scattering}},
journal={Z. Phys. A.},
year={1995},
month={Sep},
day={01},
volume={353},
number={3},
pages={273-277},
issn={0939-7922},
doi={10.1007/BF01292332},
url={https://doi.org/10.1007/BF01292332}
}

@article{gyu1975,
title = {{Higher order vacuum polarization for finite radius nuclei}},
journal = {Nuclear Physics A},
volume = {244},
number = {3},
pages = {497-525},
year = {1975},
issn = {0375-9474},
doi = {https://doi.org/10.1016/0375-9474(75)90554-0},
url = {https://www.sciencedirect.com/science/article/pii/0375947475905540},
author = {Miklos Gyulassy}
}

@manual{flint,
  key = {{FLINT}},
  author = {The {FLINT} team},
  title = {{FLINT}: {F}ast {L}ibrary for {N}umber {T}heory},
  year = {2026},
  note = {Version 3.5.0, \url{https://flintlib.org}}
}

@article{joh2017,
  author = {Fredrik Johansson},
  journal = {IEEE Trans. Comput.},
  title = {{{A}rb: Efficient Arbitrary-Precision Midpoint-Radius Interval Arithmetic}},
  year = {2017},
  volume = {66},
  number = {8},
  pages = {1281-1292},
  doi = {10.1109/TC.2017.2690633}
}

@article{yhk2025,
  title = {{One-loop electron self-energy with accelerated partial-wave expansion in the Coulomb gauge}},
  author = {Yerokhin, V. A. and Harman, Z. and Keitel, C. H.},
  journal = {Phys. Rev. A},
  volume = {111},
  issue = {1},
  pages = {012802},
  numpages = {16},
  year = {2025},
  month = {Jan},
  publisher = {American Physical Society},
  doi = {10.1103/PhysRevA.111.012802},
  url = {https://link.aps.org/doi/10.1103/PhysRevA.111.012802}
}

@article{yhk2025a,
  title = {{Two-loop electron self-energy with accelerated partial-wave expansion}},
  author = {Yerokhin, V. A. and Harman, Z. and Keitel, C. H.},
  journal = {Phys. Rev. A},
  volume = {111},
  issue = {4},
  pages = {042820},
  numpages = {12},
  year = {2025},
  month = {Apr},
  publisher = {American Physical Society},
  doi = {10.1103/PhysRevA.111.042820},
  url = {https://link.aps.org/doi/10.1103/PhysRevA.111.042820}
}

@article{mil1994,
title = {{Present status of Delbrück scattering}},
journal = {Phys. Rep.},
volume = {243},
number = {4},
pages = {183-214},
year = {1994},
issn = {0370-1573},
doi = {https://doi.org/10.1016/0370-1573(94)00058-1},
url = {https://www.sciencedirect.com/science/article/pii/0370157394000581},
author = {A.I. Milstein and M. Schumacher}
}

@article{svtg2021,
  title = {{Redefined vacuum approach and gauge-invariant subsets in two-photon-exchange diagrams for a closed-shell system with a valence electron}},
  author = {Soguel, R. N. and Volotka, A. V. and Tryapitsyna, E. V. and Glazov, D. A. and Kosheleva, V. P. and Fritzsche, S.},
  journal = {Phys. Rev. A},
  volume = {103},
  issue = {4},
  pages = {042818},
  numpages = {16},
  year = {2021},
  month = {Apr},
  publisher = {American Physical Society},
  doi = {10.1103/PhysRevA.103.042818},
  url = {https://link.aps.org/doi/10.1103/PhysRevA.103.042818}
}

@article{bctt2005,
  author       = {Beiersdorfer, P and Chen, H and Thorn, D B and Trabert, E},
  title        = {{Measurement of the two-loop Lamb shift in lithiumlike U89+}},
  url          = {https://www.osti.gov/biblio/877743},
  journal      = {Phys. Rev. Lett.},
  issn         = {ISSN PRLTAO},
  volume       = {95},
  place        = {United States},
  year         = {2005},
  month        = {04}
}

@article{bkms2003,
  title = {{Precise Determination of the $2{s}_{1/2}\mathrm{\text{\ensuremath{-}}}2{p}_{1/2}$ Splitting in Very Heavy Lithiumlike Ions Utilizing Dielectronic Recombination}},
  author = {Brandau, C. and Kozhuharov, C. and M\"uller, A. and Shi, W. and Schippers, S. and Bartsch, T. and B\"ohm, S. and B\"ohme, C. and Hoffknecht, A. and Knopp, H. and Gr\"un, N. and Scheid, W. and Steih, T. and Bosch, F. and Franzke, B. and Mokler, P. H. and Nolden, F. and Steck, M. and St\"ohlker, T. and Stachura, Z.},
  journal = {Phys. Rev. Lett.},
  volume = {91},
  issue = {7},
  pages = {073202},
  numpages = {4},
  year = {2003},
  month = {Aug},
  publisher = {American Physical Society},
  doi = {10.1103/PhysRevLett.91.073202},
  url = {https://link.aps.org/doi/10.1103/PhysRevLett.91.073202}
}

@article{bbkc2007,
  title = {{Testing QED Screening and Two-Loop Contributions with He-Like Ions}},
  author = {Bruhns, H. and Braun, J. and Kubi\ifmmode \check{c}\else \v{c}\fi{}ek, K. and Crespo L\'opez-Urrutia, J. R. and Ullrich, J.},
  journal = {Phys. Rev. Lett.},
  volume = {99},
  issue = {11},
  pages = {113001},
  numpages = {4},
  year = {2007},
  month = {Sep},
  publisher = {American Physical Society},
  doi = {10.1103/PhysRevLett.99.113001},
  url = {https://link.aps.org/doi/10.1103/PhysRevLett.99.113001}
}

@article{bss2021,
  title={{Perspectives on testing fundamental physics with highly charged ions in Penning traps}},
  author={Blaum, Klaus and Eliseev, Sergey and Sturm, Sven},
  journal={Quantum Science \& Technology},
  volume={6},
  number={1},
  pages={014002},
  year={2021},
  publisher={IOP Publishing}
}

@article{kscs2018,
  title = {{Highly charged ions: Optical clocks and applications in fundamental physics}},
  author = {Kozlov, M. G. and Safronova, M. S. and Crespo L\'opez-Urrutia, J. R. and Schmidt, P. O.},
  journal = {Rev. Mod. Phys.},
  volume = {90},
  issue = {4},
  pages = {045005},
  numpages = {49},
  year = {2018},
  month = {Dec},
  publisher = {American Physical Society},
  doi = {10.1103/RevModPhys.90.045005},
  url = {https://link.aps.org/doi/10.1103/RevModPhys.90.045005}
}

@article{aam2013,
title = {{Table of experimental nuclear ground state charge radii: An update}},
journal = {At. Data Nucl. Data Tables},
volume = {99},
number = {1},
pages = {69-95},
year = {2013},
issn = {0092-640X},
doi = {https://doi.org/10.1016/j.adt.2011.12.006},
url = {https://www.sciencedirect.com/science/article/pii/S0092640X12000265},
author = {I. Angeli and K.P. Marinova}
}

@misc{mal2023,
  author = {A. Malyshev},
  howpublished = {private communication}
}

@article{mgka2021,
  title = {{Ab initio Calculations of Energy Levels in Be-Like Xenon: Strong Interference between Electron-Correlation and QED Effects}},
  author = {Malyshev, A. V. and Glazov, D. A. and Kozhedub, Y. S. and Anisimova, I. S. and Kaygorodov, M. Y. and Shabaev, V. M. and Tupitsyn, I. I.},
  journal = {Phys. Rev. Lett.},
  volume = {126},
  issue = {18},
  pages = {183001},
  numpages = {9},
  year = {2021},
  month = {May},
  publisher = {American Physical Society},
  doi = {10.1103/PhysRevLett.126.183001},
  url = {https://link.aps.org/doi/10.1103/PhysRevLett.126.183001}
}

@article{jas1985,
title = {{The lamb shift in hydrogen-like atoms, 1 $\leq$ Z $\leq$ 110}},
journal = {At. Data Nucl. Data Tables},
volume = {33},
number = {3},
pages = {405-446},
year = {1985},
issn = {0092-640X},
doi = {https://doi.org/10.1016/0092-640X(85)90010-5},
url = {https://www.sciencedirect.com/science/article/pii/0092640X85900105},
author = {W.R. Johnson and Gerhard Soff}
}

@article{ppv2017,
  title = {{Testing fundamental interactions on the helium atom}},
  author = {Pachucki, Krzysztof and Patk\'o\ifmmode \check{s}\else \v{s}\fi{}, Vojt\ifmmode \check{e}\else \v{e}\fi{}ch and Yerokhin, Vladimir A.},
  journal = {Phys. Rev. A},
  volume = {95},
  issue = {6},
  pages = {062510},
  numpages = {8},
  year = {2017},
  month = {Jun},
  publisher = {American Physical Society},
  doi = {10.1103/PhysRevA.95.062510},
  url = {https://link.aps.org/doi/10.1103/PhysRevA.95.062510}
}

@article{sap1998,
  title = {{Theoretical methods for the relativistic atomic many-body problem}},
  author = {Sapirstein, J.},
  journal = {Rev. Mod. Phys.},
  volume = {70},
  issue = {1},
  pages = {55--76},
  numpages = {0},
  year = {1998},
  month = {Jan},
  publisher = {American Physical Society},
  doi = {10.1103/RevModPhys.70.55},
  url = {https://link.aps.org/doi/10.1103/RevModPhys.70.55}
}

\end{document}